\documentstyle[12pt,epsfig,amsmath,amssymb,amsbsy]{article}
\def\tp{t_\perp}
\def\est{E^*}
\def\kp{k_{\perp}}
\def\ep{\epsilon_{\perp}}
\def\figwidth{0.5\textwidth}

\begin{document}
\pagestyle{empty}

\fontfamily{ptm}
\selectfont

T. Giamarchi, S. Biermann, A. Georges, A. Lichtenstein

\vspace{5.5cm}

\noindent \hfill  \parbox{16.5cm}{ \footnotesize {\bf Abstract.}
The Bechgaard salts are made of weakly coupled one dimensional
chains. This particular structure gives the possibility to observe
in these systems a dimensional crossover between a high
temperature (or high energy) one dimensional phase  and a two or
three dimensional system. Since the filling of the chains is
commensurate the system thus undergoes a deconfinement transition
from a one dimensional Mott insulator to a two (or three)
dimensional metal. Such a transition has of course a strong impact
on the physical properties of these compounds, and is directly
seen in transport measurements.

In order to describe such a transition a dynamical mean field
method has been introduced (chain-DMFT). Using this method we
investigate a system of coupled Hubbard chains and show that we
can indeed reproduce the deconfinement transition. This allows to
determine physical quantities such as the transport transverse to
the chains and the shape of the Fermi surface and quasiparticle
residues in the low temperature phase.

\vspace{.5cm}

}

\vspace{.9cm}

\noindent
{\bf 1. INTRODUCTION}
\vspace{.4cm}

\noindent The Bechgaard salts TMTSF$_2$X were the first organic
compounds to exhibit superconductivity, and have thus been the
focus of intense experimental and theoretical studies
\cite{jerome_organic_review,bourbonnais_jerome_review}. These
systems are highly anisotropic crystals that consist of parallel
conducting chains. The electron transfer integrals along the
chains (in the $a$ direction) and transverse to the chains (in the
$b$ and $c$ directions) are typically $t_a = 3000$ K, $t_b = 300$
K, and $t_c = 10$ K. Therefore one can think of these materials as
one-dimensional chains coupled by small inter-chain hoppings.
Given the hierarchy of transverse coupling the system is first
expected to become two dimensional and then three dimensional at
low temperatures.

Despite the fact that they have been studied for more than twenty
years, these compounds are still largely not understood. Some of
the most important open questions are: (i) the nature and
properties of the the high temperature ``normal'' phase. (ii) what
happens when the energy (temperature, frequency, etc.) is lowered
and one can expect the various hopping integrals to play a role.
We will briefly review in this paper the recent progress that have
been made to understand these two questions. \vspace{.6cm}

\noindent
{\bf 2. ``Normal'' state}
\vspace{.4cm}

\noindent Since the interchain hopping are so widely separated one
would expect that when the temperature or energy (frequency, etc.)
is higher than $t_b$ the interchain hopping cannot occur
coherently and the systems are in a one dimensional limit. In that
case one would expect due to interactions a Luttinger liquid
behavior \cite{giamarchi_1dbook}. However, although the Bechgaard
salts themselves are metallic, the parent compounds TMTTF (Fabre
salts) show a marked insulating behavior. Such a behavior is
obviously due to interactions and related to Mott physics. Indeed,
by chemistry both TMTTF and TMTSF compounds are quarter filled.
Such insulating behavior is thus the proof of the importance of
interactions in TMTTF, even at relatively high energies. For the
TMTSF the question is more subtle in view of the metallic behavior
at ambient pressure and whether such compound is a Luttinger
liquid or a Fermi liquid was the subject of considerable debate
\cite{gorkov_sdw_tmtsf}. Another important question is of course
the reason for such a difference between the very close families
TMTTF and TMTSF, for which the various characteristics (bandwidth,
dimerization, interactions) vary relatively little.

One possible interpretation of this difference was suggested a
long time ago \cite{emery_umklapp_dimerization} and attributed to
the dimerization of the chains. Indeed because of the existence of
a slight dimerization, the effective band is half filled and not
quarter filled and one is now in the standard situation to get a
Mott insulator (one particle per site). In such a case it is well
known that in one dimension any repulsive interaction leads to a
Mott insulating state. The accepted idea was thus that in the
TMTSF family, dimerization being much smaller, the system would be
viewed as quarter filled and thus could not be a Mott insulator.
The commonly
adopted
point of view to describe these systems would thus be
to describe them by a Hubbard like model, where the insulating
character would mostly depend on the local interaction $U$ and on
the dimerization $\Delta_d$.

However this point of view had to be seriously reexamined in the
recent years. On the experimental side, optical data
\cite{dressel_optical_tmtsf,schwartz_electrodynamics} shows quite
clearly that even the TMTSF family exhibits strong interaction
effects. Indeed, although such compound seems to have a rather
``standard'' d.c. conductivity, all the d.c. transport is in fact
due to a very narrow Drude peak containing only 1\% of the
spectral weight, whereas 99\% of the spectral weight is above an
energy gap (of the order of $200$ cm$^{-1}$), and is reminiscent
of a Mott insulating structure. On the theory side, it was shown
that the Mott insulating state could exist for higher
commensurabilities than one particle per site
\cite{giamarchi_curvature,schulz_mott_revue} for strong enough and
{\it non on-site} interactions. Physical quantities such as the
optical conductivity where computed for such one dimensional Mott
insulators \cite{giamarchi_umklapp_1d,giamarchi_mott_shortrev}.
Since even purely quarter filled systems could be Mott insulators,
this prompted for the question of which mechanism was dominant for
the gap observed in the TMTSF optical data: (i) the ``half
filled'' nature of the band, due to the weak dimerization of the
band; (ii) the direct mechanism responsible for the Mott
insulating behavior of a quarter filled interacting system (no
dimerization is needed).

Let us examine the two mechanisms (for more details see
\cite{giamarchi_mott_shortrev,schwartz_electrodynamics}). In order
to get an insulator in one dimension one has to consider the
interaction processes that do not conserve momentum (so called
Umklapp processes). For a system at half filling the umklapp takes
two electrons from the left part of the Fermi surface (i.e. with
momenta close to $-k_F$) and transfers them to the right part of
the Fermi surface (i.e. with momenta close to $+k_F$). The total
momentum transferred is thus $4 k_F$ and for a half filled band
this is $2 \pi$, i.e. a vector of the reciprocal lattice. For a
half filled system any interaction process is thus able to lead to
an umklapp process, and thus the strength of the umklapp is of
order $U$ if $U$ is the interaction. If the band is quarter filled
such a process is inefficient since the transferred momentum would
be $4 k_F = \pi$, and thus different from a vector of the
reciprocal lattice. However in the presence of a slight
dimerization of the chains one should diagonalize the kinetic
energy and the eigenstates of the system are now mixtures of
states containing $k$ and $k+\pi$. In the presence of a
dimerization $\Delta_d$ it is thus easy to show that an umklapp
process should exist and that its strength is given by $g_{3,1/2}
\sim U \Delta_d/W$ where $W$ is the bandwidth of the system. Thus
if the dimerization is zero one recovers that no umklapp process
corresponding to half filling exists. In one dimension such an
umklapp is relevant for any repulsive interactions and would open
a gap in the spectrum of order $\Delta_{1/2} \sim W
(g_{3,1/2}/W)^{1/(2-2K_\rho)}$, where $K_\rho$ is the so-called
Luttinger liquid parameter. $K_\rho < 1$ corresponds to repulsive
interactions. $g_{3,1/2}$ is thus relevant for $K_\rho < 1$.
However an umklapp process can also exists directly for a quarter
filled system, if one transfers {\it four} electrons from one side
of the Fermi surface to the other. Since an interaction process
only scatters two electrons, such a process can only be generated
at the third order in perturbation theory. The corresponding
umklapp strength is of order $g_{3,1/4} \sim U (U/W)^2$. This
process is only relevant if the Luttinger parameter is smaller
than $K_\rho < 1/4$. Note that to obtain such a value it is
necessary in addition to having strong interactions to have finite
range (i.e. more than on site) interactions. Indeed for the
Hubbard model the smallest possible value of the Luttinger
parameter is $K_\rho > 1/2$.

Thus in the organic compounds, both process are presents and the
question is whether one of the two is dominant
\cite{giamarchi_mott_shortrev}. Clearly this depends on the
interactions: if the interactions are weak then the $g_{3,1/4}$
can obviously be neglected. Not only the coupling constant is
small since it contains the factor $(U/W)^2$ but the process is
also irrelevant. On the other hand, if the interactions are strong
(i.e. of the order of the bandwidth), then $g_{3,1/4}/W \sim 1$
whereas $g_{3,1/2}/W \sim \Delta_d/W$ which is quite small (a few
percents). Estimates of the interactions in the TM family suggests
rather strong interactions, a sign that the system could be
dominated by quarter filled umklapps. A clear answer to this
question is provided by the optical conductivity
\cite{schwartz_electrodynamics}. For a Mott insulator in one
dimension, the behavior of the conductivity above the Mott gap
should be $\sigma(\omega) \propto \omega^{4 K_\rho - 5}$ for half
filling and $\sigma(\omega) \propto \omega^{16 K_\rho - 5}$ for
quarter filling. The observed behavior fits well a power-law
$\omega^{-1.3}$. If one tries to fit this behavior to an half
filled process this leads to exceedingly weak interactions $K \sim
1 - U/(\pi v_F) \sim 0.93$. Using the expression for the Mott gap
would lead to a gap which is orders of magnitude smaller than the
observed optical gap $\Delta \sim 200$cm$^{-1}$. Thus the only
possibility to fit the optical data is to assume that for the
TMTSF family the conductivity is in fact dominated by the quarter
filled umklapp. The value of $K_\rho$ extracted from the optics
$K_\rho \sim 0.23$ corresponds indeed to quite strong interactions
and thus leads to a consistent description.

This prompts for several conclusions: (i) In order to modelize the
TMTSF it is thus necessary to include in addition to a local
interaction $U$ a quite strong nearest neighbor (at least -- and
possibly longer range) repulsion $V$. Although one can show the
importance of this term for TMTSF, given the similarities between
the two systems such a term need also to be taken into account for
TMTTF also. (ii) Given the proximity of $K =0.23$ to the value at
which the quarter filled umklapp is irrelevant, the observed
decrease of the gap under pressure could be attributed to the
proximity of this quantum critical point. Indeed applying pressure
increases the hoppings, therefore reducing the interactions
compared to the bandwidth and making $K_\rho$ larger. One can thus
expect the gap to vary rapidly with pressure. (ii) Although
dimerization is clearly present and certainly plays a role in some
of the physical quantities, most of the physics seems to be
directly controlled by the quarter filled nature of the band. This
strongly suggests that most of the physics of the TMTSF family
should also be observed in a purely quarter filled compound.
Compounds without dimerization have been synthesized and have been
showed to be (quarter-filled) insulators (see e.g.
\cite{heuze_quarterfilled_refs}). The existence of such insulators
confirms the above description. It would be extremely interesting
to carefully study these systems under pressure, since they should
exhibit most of the properties of the Bechgaard salts, including
most probably the superconductivity.

\vspace{.6cm}

\noindent {\bf 3. Deconfinement transition} \vspace{.4cm}

\noindent Let us now move to the effects of interchain hopping.
Due to the interchain hopping a dimensional crossover will take
place at low energy between decoupled chains and a higher-
dimensional behavior. Since the isolated chains would be
insulators, the interchain hopping can induce a deconfinement
transition provided that it becomes larger than the Mott gap. The
system will thus crossover from a regime where one has essentially
uncoupled (insulating) chains to that of metallic planes. The
interpretation that the change of behavior between the insulating
and metallic regimes is indeed due to such deconfinement
transition \cite{giamarchi_mott_shortrev} can be strengthened by
the optical data. A measure of the gap extracted from the optical
conductivity shows that the change of nature occurs when the
observed gap is roughly of the order of magnitude of the
interchain hopping \cite{vescoli_confinement_science}.
Understanding the characteristics of such a transition (energy
scale, critical values of the hopping, physical nature of the
various phases) is one of the most challenging questions on these
systems, on which we shall focus in the following (for more
details see
\cite{biermann_dmft1d_hubbard_short,biermann_oned_crossover_review}).

Perturbation theory in the interchain hopping $\tp$ is only of
limited use for the study of the crossovers. For incommensurate
fillings a perturbative RG treatment indicates that this effective
inter-chain hopping grows as the energy scale is reduced: the
crossover will occur when this running coupling reaches $kT$.
While this perturbative RG analysis allows to estimate a scale
$\est$
for the dimensional crossover
\cite{bourbonnais_rmn,brazovskii_transhop}, it breaks down for
$T<\est$ since the effective $\tp$ flows to large values. In
particular, it does not provide information on the detailed nature
of the low-T Fermi liquid regime. Thus, a proper handling of the
dimensional crossover in quasi one-dimensional systems has to
resort to techniques which are not perturbative in $\tp$. This is
even clearer in the case of a commensurate filling. If one starts
from the 1D Mott insulator fixed point, the deconfinement
transition is clearly a non-perturbative phenomenon since the
inter-chain hopping is an irrelevant perturbation at this fixed
point. If, on the other hand, one starts with the LL fixed point
associated with the high-T regime, then one has to deal
simultaneously with {\it two} relevant perturbations: the umklapp
scattering (responsible for Mott physics) and the interchain
hopping. Non-perturbative studies are thus needed to investigate
both the deconfinement transition and the dimensional crossover.
Recently we have developped a mean field approach designed to
handle {\it an infinite array of coupled chains} in a
non-perturbative manner
\cite{arrigoni_tperp_resummation,georges_organics_dinfiplusone,biermann_dmft1d_hubbard_short}.
The details of the method
\cite{biermann_dmft1d_hubbard_short,biermann_oned_crossover_review}
can be found elsewhere and we focus here on some of the results
that could be extracted.

An analytical solution of the mean field equations is
impossible. Thus only few quantities such as the Drude weight of
the transverse conductivity could be obtained analytically
\cite{georges_organics_dinfiplusone}. In order to see whether our
mean field approximation was able to capture the deconfinement
transition we used it on a simpler model, namely a model of
coupled Hubbard chains at half filling. Let us point out that this
is only a toy model, the proper model for the organics being a
$U,V$ model at quarter filling. Nevertheless, as far as the
deconfinement transition is concerned, this toy model should
contain most of the necessary ingredients and is simpler to
simulate, in order to test the method. More complicated
calculation on the more realistic model being necessary as a
second step to confirm the results of the toy model.

\begin{figure}
\centerline{\includegraphics[width=\figwidth]{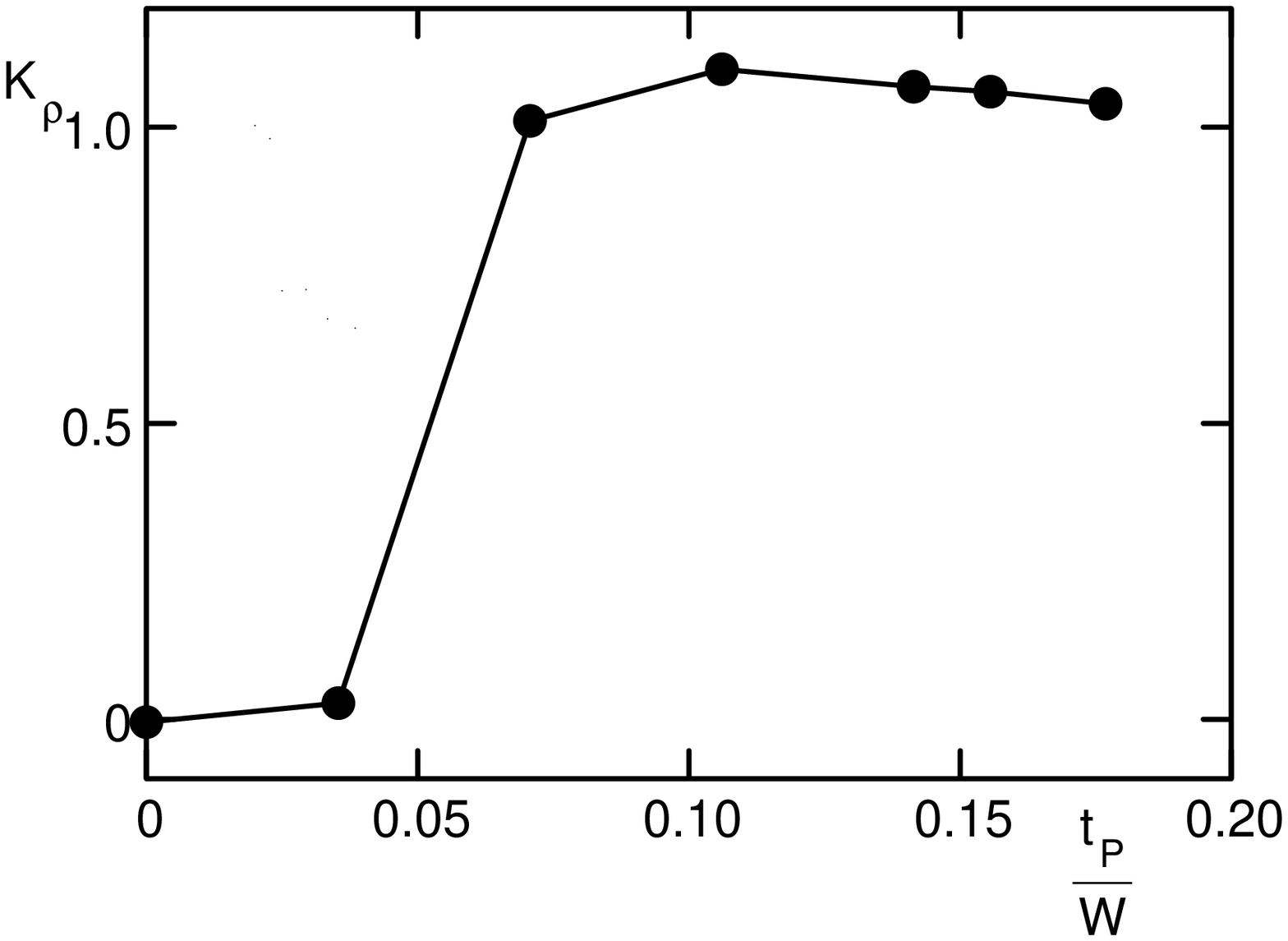}
             \includegraphics[width=\figwidth]{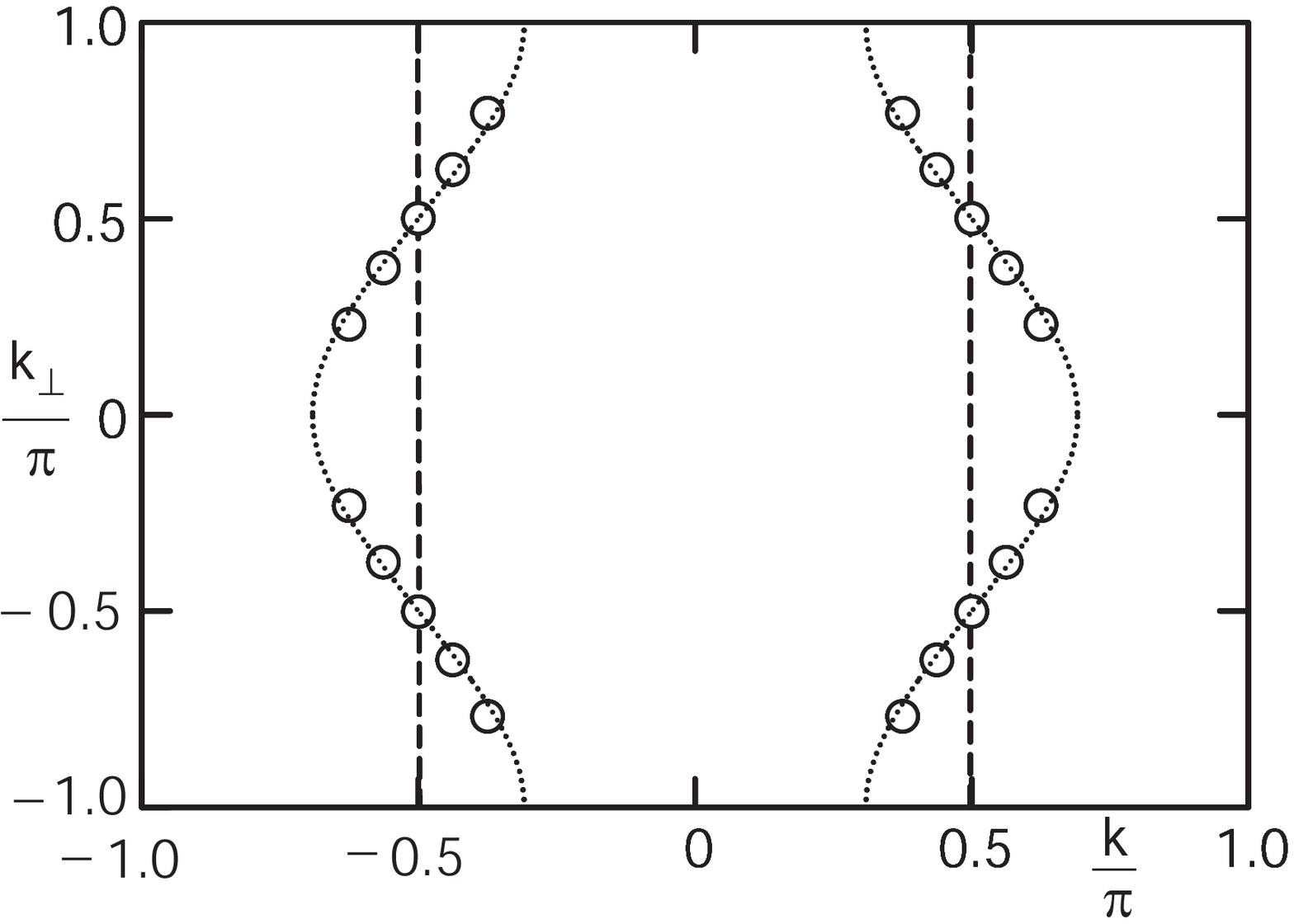}}
\caption{(left) $K_\rho$ as a function of the perpendicular
hopping $t_\perp$ for a half filled system. The deconfinement
transition is clearly visible. (right) Fermi surface (FS) in the
half-filled case with $\tp/W=0.14$, $U/W=0.65$ (circles), compared
to the FS of the non-interacting case (dotted line) and of the
purely 1d case ($\tp=0$ -dashed-.). After {\protect
\cite{biermann_dmft1d_hubbard_short,biermann_oned_crossover_review}}.}
\label{fig:un}
\end{figure}

In Fig.\ref{fig:un}, we display the effective $K_\rho$ as a
function of interchain hopping, for $U/W=0.65$ and at a rather low
temperature $T/W=0.025$.
For small $\tp/W$, the value $K_\rho=0$ indicates a Mott
insulating behavior (with a decay of the spin-spin correlation
similar to that of a Heisenberg spin chain). In that regime, the
calculated charge correlation function (not shown) clearly
displays the exponential decay associated with a finite charge
gap. Beyond a critical value of $\tp/W$, we find $K_\rho\simeq 1$,
signalling a FL regime (and a corresponding behavior for the
charge correlation function). Hence, the expected deconfinement
transition is clearly revealed by our calculations. The location
of the deconfinement transition is in reasonable agreement with
the naive criterion $\Delta_{1D} \sim t_\perp^{\rm eff}$, with
$\tp^{\rm eff}$ the renormalized inter-chain hopping.
\cite{giamarchi_mott_shortrev,vescoli_confinement_science,%
tsuchiizu_confinement_spinful_refs}. At strong enough transverse
coupling the system becomes a FL. In our numerical simulations,
the onset of the FL regime is identified from the behavior of
$K_\rho$ (see Fig.~\ref{fig:un}) and from a linear behavior of the
imaginary part of the self-energy in Matsubara space: $\Sigma(k, i
\omega) \sim i \omega$. The equation defining the Fermi surface
$\mu-\epsilon_k-\Sigma(k,0)-\ep=0$ then yields a relation $\kp(k)$
for the points $(k,\kp)$ that lie on the Fermi surface. These are
visualized in Fig.~\ref{fig:un} for the half-filled case. For the
uncoupled (1d) system the Fermi surface consists of straight lines
(dashed lines in the figure); the transverse hopping induces some
cosine-like modulation but does not change the topology
drastically. Indeed, the Fermi surface of the interacting coupled
system (circles in Fig.~\ref{fig:un}) is very close to the one of
non-interacting ($U=0$.) coupled chains (dotted line in
Fig.~\ref{fig:un}).

\begin{figure}
 \centerline{\includegraphics[width=\figwidth]{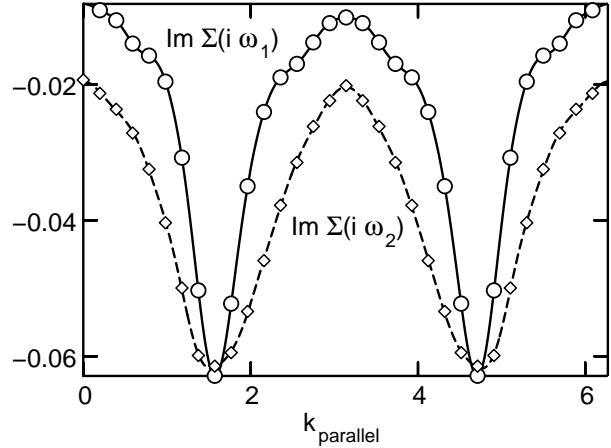}}
 \caption{Imaginary part of the self energy as a function of $k$ along the chains for
 $\beta=40$, $U=0.65$, $t\perp=0.05/\sqrt{2}$.
$\omega_1= \pi/\beta$ denotes the lowest Matsubara frequency
and $\omega_2 = 3\pi/\beta$ the second one.
}
\label{figdeux}
\end{figure}

A very interesting question, is whether the strong interactions
can lead to the existence of ``hot spots'' (see e.g.
\cite{zheleznyak_hot_spots,parcollet_hotspots_dmft}) on the Fermi
surface of the system at low energy. Clearly this question needs
to be investigated further but our calculation can bring some
answers to this question. If one looks inside the 3D metallic
regime ($K_\rho = 1$) and computes the QP residue $Z_{\kp}$ then
only a very small dependence on the Fermi surface points is seen
(Table I), with very shallow minima at $\kp\sim \pm \pi/2$. This
small variation is however on the scale of our error bars. .
\begin{table}
\caption{QP weights $Z(\kp)$ for different points on the
FS (half-filled case, $\tp=0.14 W$, $U/W=0.65$).
\label{tab1half}}
\begin{center}
\begin{tabular}{llllll}
$\kp/\pi$ & 0.23 & 0.38 & 0.50 & 0.62 & 0.77 \\
\hline
$Z(\kp)$  & 0.79  & 0.77 & 0.76 & 0.77 & 0.79
\end{tabular}
\end{center}
\end{table}
This is true even for the closest point to the transition.
Therefore in the metallic regime we do not find signs of ``hot spots''
within our approximation. On the other hand the situation is quite
different if one is close to the transition but on the $K_\rho =
0$ side (see Fig.~\ref{figdeux})
In that case the imaginary part of the self energy shows a strong
dip. In the insulator this dip would signal the divergence $\Sigma
\sim \Delta/(i \omega_n)$ for $\beta \to \infty$
and $k_{\parallel}=\pm \pi/2$ characteristic of an insulator. At
finite temperature, the ratio of the imaginary part of the
self-energy for the two lowest Matsubara frequencies would be
$\sim \omega_2/\omega_1 = 3/2$. However close to the transition
this ratio is much closer to one, as can be seen on the figure.
Whether the fact that it seems to be still greater than one is
simply a finite size artefact, and the system is in reality in an
intermediate metallic regime with very small quasiparticle residue
around the points $k_\parallel = \pm \pi/2$ is an interesting open
issue that clearly deserves further investigation.

\vspace{.6cm}

\noindent {Acknowledgements} \vspace{.4cm}

We acknowledge many discussions with D. Jerome, P. Auban-Senzier,
L. Degiorgi, G. Gr{\"u}ner and O. Parcollet. This work is partly
supported by the Swiss national fund under MaNEP and a grant of
supercomputing time at IDRIS Orsay (project number 031393).

\noindent

\begin{thebibliography}{10}

\bibitem{jerome_organic_review}
D. J{\'e}rome,  in {\em Organic Superconductors: From
{(TMTSF)$_2$PF$_6$} to
  Fullerenes} (Marcel Dekker, New York, 1994), Chap.~10, p.\ 405.

\bibitem{bourbonnais_jerome_review}
C. Bourbonnais and D. J{\'e}rome,  in {\em Advances in Synthetic
Metals, Twenty
  years of Progress in Science and Technology}, edited by P. Bernier, S.
  Lefrant, and G. Bidan (Elsevier, New York, 1999), p.\ 206, preprint
  cond-mat/9903101.

\bibitem{giamarchi_1dbook}
T. Giamarchi, {\em Quantum Physics in One Dimension} (Clarendon
Press, Oxford,
  2003).

\bibitem{gorkov_sdw_tmtsf}
L.~P. Gorkov, Physica B {\bf 230-232},  970  (1997).

\bibitem{emery_umklapp_dimerization}
V.~J. Emery, R. Bruinsma, and S. Barisic, Phys. Rev. Lett. {\bf
48},  1039
  (1982).

\bibitem{dressel_optical_tmtsf}
M. Dressel, A. Schwartz, G. Gr{\"u}ner, and L. Degiorgi, Phys.
Rev. Lett. {\bf
  77},  398  (1996).

\bibitem{schwartz_electrodynamics}
A. Schwartz {\it et~al.}, Phys. Rev. B {\bf 58},  1261  (1998).

\bibitem{giamarchi_curvature}
T. Giamarchi and A.~J. Millis, Phys. Rev. B {\bf 46},  9325
(1992).

\bibitem{schulz_mott_revue}
H.~J. Schulz,  in {\em Strongly Correlated Electronic Materials:
The Los Alamos
  Symposium 1993}, edited by K.~S. {Bedell {\it et al.}} (Addison--Wesley,
  Reading, MA, 1994), p.\ 187.

\bibitem{giamarchi_umklapp_1d}
T. Giamarchi, Phys. Rev. B {\bf 44},  2905  (1991).

\bibitem{giamarchi_mott_shortrev}
T. Giamarchi, Physica B {\bf 230-232},  975  (1997).

\bibitem{heuze_quarterfilled_refs}
K. Heuz{\'e} {\it et~al.}, Adv. Mater. {\bf 15},  1251  (2003),
and references
  therein.

\bibitem{vescoli_confinement_science}
V. Vescoli {\it et~al.}, Science {\bf 281},  1191  (1998).

\bibitem{biermann_dmft1d_hubbard_short}
S. Biermann, A. Georges, A. Lichtenstein, and T. Giamarchi, Phys.
Rev. Lett.
  {\bf 87},  276405  (2001).

\bibitem{biermann_oned_crossover_review}
S. Biermann, A. Georges, T. Giamarchi, and A. Lichtenstein,  in
{\em Strongly
  Correlated Fermions and Bosons in Low Dimensional Disordered Systems}, edited
  by I.~V. {Lerner {\it et al.}} (Kluwer Academic Publishers, Dordrecht, 2002),
  p.\ 81, cond-mat/0201542.

\bibitem{bourbonnais_rmn}
C. Bourbonnais {\it et~al.}, J. Phys. (Paris) Lett. {\bf 45},
L755  (1984).

\bibitem{brazovskii_transhop}
S. Brazovskii and V. Yakovenko, J. Phys. (Paris) Lett. {\bf 46},
L111  (1985).

\bibitem{arrigoni_tperp_resummation}
E. Arrigoni, Phys. Rev. Lett. {\bf 83},  128  (1999).

\bibitem{georges_organics_dinfiplusone}
A. Georges, T. Giamarchi, and N. Sandler, Phys. Rev. B {\bf 61},
16393
  (2000).

\bibitem{tsuchiizu_confinement_spinful_refs}
M. Tsuchiizu, P. Donohue, Y. Suzumura, and T. Giamarchi, Eur.
Phys. J. B {\bf
  19},  185  (2001), and references therein.

\bibitem{zheleznyak_hot_spots}
A.~T. Zheleznyak and V.~M. Yakovenko, Synth. Metal {\bf 70},  1005
(1995).

\bibitem{parcollet_hotspots_dmft}
O. Parcollet, G. Biroli, and G. Kotliar, cond-mat/0308577, 2003.

\end{thebibliography}

\newpage

\noindent {\Large \bf Dimensional crossover and deconfinement in
Bechgaard salts} \vspace{1cm}

\noindent \underline{T. Giamarchi}$^a$,  S. Biermann$^b$, A.
Georges$^c$,  and  A. Lichtenstein$^d$ \vspace{1cm}

\noindent
$^a${\it University of Geneva, 24 Quai Ernest Ansermet,
1211
Geneva, Switzerland \\
e-mail:thierry.giamarchi@physics.unige.ch,}  \\
$^b${\it LPTENS-CNRS
UMR 8549, 24 Rue Lhomond 75231 Paris
Cedex 05, France \\
e-mail: Silke.Biermann@cpht.polytechnique.fr} \\
$^c${\it LPTENS-CNRS UMR 8549, 24 Rue Lhomond 75231 Paris Cedex
05, France \\ e-mail: georges@lpt.ens.fr} \\
$^d${\it University of Nijmegen, NL-6525 ED Nijmegen, The
Netherlands \\
e-mail: A.Lichtenstein@sci.kun.nl}

\end{document}